\documentclass[prl,twocolumn,superscriptaddress,showkeys,amsmath,amsfonts,
               floatfix]{revtex4}

\usepackage{epsfig,psfrag}
\usepackage{CJK}

\usepackage{graphicx}
\usepackage{dcolumn}
\usepackage{bm}
\usepackage{amssymb}
\usepackage{natbib}
\usepackage{upgreek}

\begin{document}
\title{Optical absorption of angulon in metal halide perovskites}
\author{Jia-Wei Wu}
\affiliation{Tianjin Key Laboratory of Low Dimensional Materials Physics and Preparing Technology,
Department of Physics, School of science, Tianjin University, Tianjin 300354, China}
\author{Yu Cui}
\email{cuiyu@tju.edu.cn}
\affiliation{Tianjin Key Laboratory of Low Dimensional Materials Physics and Preparing Technology,
Department of Physics, School of science, Tianjin University, Tianjin 300354, China}
\author{Shao-Juan Li}
\affiliation{State Key Laboratory of Luminescence and Applications, Changchun Institute of Optics, Fine Mechanics and Physics Chinese Academy of Sciences, Changchun 130033, China}
\author{Zi-Wu Wang}
\email{wangziwu@tju.edu.cn}
\affiliation{Tianjin Key Laboratory of Low Dimensional Materials Physics and Preparing Technology,
Department of Physics, School of science, Tianjin University, Tianjin 300354, China}

\begin{abstract}
We theoretically study the optical absorption of an angulon in the metal halide perovskites (MHP) based on the improved Devreese-Huybrechts-Lemmens model, where the formation of quasiparticle angulon states originates from the organic cation rotating in the inorganic octahedral cage of MHP. We find that the resonance optical absorption peaks are appeared when the energy of incident photon matches the quantum levels of angulon. Moreover, the intensity of absorption depends on the quantum states of phonon angular momentum. These theoretical results provide significant insight to study the redistribution of angular momenta for the rotational molecules immersed into the many-body environment.
\end{abstract}
\maketitle

\section{Introduction}
 Angulon describing the entity of the rotational particles (molecule or impurity)  dressed by the phononic bath around it, has been proposed by Schmidt and Lemeshko for the first time in 2015\cite{wjw1}. With the help of this quasiparticle concept, the intractable angular momentum algebra for the rotational molecules (or impurities) immersed into a many-body environment is tremendously simplified without the accuracy of calculation\cite{wjw2,wjw3,wjw4,wjw5,wjw6,wjw7,wjwz1,wjwj1,wjw8,wjw9,wjw10,wjw11}. They predicted a series of novel properties, such as the rotational fine structure of angulon\cite{wjw1,wjw2,wjw8}, the angular self-localization effect\cite{wjw7,wjwz1,wjw8} and the possible realization of magnetic monopoles\cite{wjw4,wjw9}, for molecules rotating in the systems of superfluid helium nanodroplets and Bose-Einstein Condensates\cite{wjw5,wjw6,wjw10}. Some of these properties have been proved by the subsequent experiments\cite{wjw5,wjwz2,wjw12,wjw13,wjwj2}. More importantly, this qusiparticle model can be expanded to study the general systems consisting of molecules rotating in the many-body bath and open up a new door to explore the angular momentum exchange between the rotational particles and its surrounding environment\cite{wjw10}.

Metal halide perovskites (MHP) with the general formula of ABX$_3$, (A, B and X represent the monovalent organic or inorganic cation, the divalent Pb$^{2+}$ cation and the halide anion, respectively) have aroused extensively interesting owing to their potential applications in optoelectronic and photovoltaic devices\cite{wjw14,wjw15}. One of their unique features is A cation rotating in the BX$_6^{4-}$ cage as shown in Fig. 1, which plays a vital role in modifying the fundamental properties of MHP\cite{wjw16,wjw17,wjw18}. Some recent experiments have proved that the reorientation of A cation could be thermally activated at certain temperature with the residence time varying from femtoseconds to picoseconds\cite{wjw18, wjw19}. Moreover, its coupling with the surrounding inorganic framework may give rise to crystal phase transitions between the orthorhombic, tetragonal, and cubic phases with temperature\cite{wjw20,wjw21}. Zhu et al. also certified that the reorientational motion of dipolar A cations could be thermally activated and their coupling with BX$_6^{4-}$ octahedral cage may affect the dynamics of charge carriers\cite{wjw22}. In particular, the ultrafast rotational motion of the A cation in response to the photo-generation of charge carriers induces the formation of large polarons\cite{wjw23,wjw24}, which results in an effective screening of the Coulomb potential to hinder hot-carrier scattering. Although the rotational motion of the A cation in MHP has been extensively investigated, studies regarding the angular momentum exchange of the rotational A cation with the BX$_6^{4-}$ octahedral cage and photon, to our best knowledge, are still lacking until now.
\begin{figure}
\includegraphics[width=3.2in,keepaspectratio]{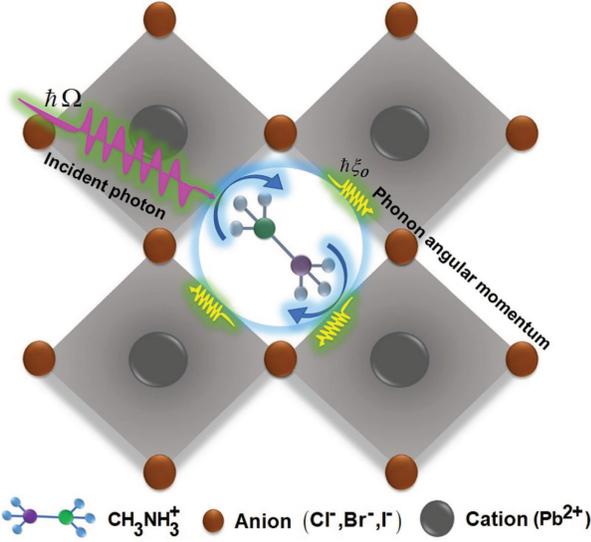}
\caption{\label{compare} (a) Schematic diagrams of an angulon and its optical absorption process, in which an angulon is formed arising from a CH$_3$NH$_3^{+}$ cation rotating in the BX$_6^{4-}$ octahedral cage. The resonance optical absorption will be occurred when the incident photon momentum is just equal to the eigenenergy of angulon consisting of the rotating CH$_3$NH$_3^{+}$ cation dressed by phonon angular momenta.}
\end{figure}

In the present paper, we employ MHP as a new exploring platform for angulon. We focus on the optical absorption process of an angulon schemed in Fig. 1, in which an angulon is formed arising from a CH$_3$NH$^+_3$ cation rotating in the BX$_6^{4-}$ octahedral cage. The resonance optical absorption coefficients are obtained based on the improved Devreese-Huybrechts-Lemmens (DHL) model that used to study the optical absorption of the traditional polaron extensively in the past decades\cite{wjw25,wjw26,wjw27,wjw28}. We compare the intensities of the optical absorption of angulon for different phonon angular momentum quantum states. These theoretical results not only enrich the knowledge of the rotational molecules immersed in the many-body systems, but also provide a simple and direct method to identify the angulon quasiparticle and phonon angular momenta around it.

\section{Two canonical transformations}
In the frame of the angulon model, the total Hamiltonian for a molecular cation rotating in the octahedral cage schemed in Fig. 1, can be written as\cite{wjw1,wjw2,wjw5,wjw8}:
\begin{equation}
H = {H_R} + {H_{ph}} + {H_{R-ph}},
\end{equation}
with
\[\begin{array}{l}
\\
{H_R} = \hbar\xi_0 \hat {\bf{L}}^{2},\\
\\
{H_{ph}} = \sum\limits_{k \lambda \mu} {\hbar {\omega_{\upsilon}}a_{k \lambda \mu}^\dag {a_{k \lambda \mu}}} ,\\
\\
{H_{R-ph}} =\sum\limits_{k \lambda \mu} {{U_{\lambda}(k)}\left[ {\hat{Y}^{*}_{\lambda \mu}(\hat{\theta},\hat{\phi}) a_{k \lambda \mu}^\dag + \hat{Y}_{\lambda \mu}(\hat{\theta},\hat{\phi}){a_{k \lambda \mu}}} \right]},\\
\end{array}\]
where the first term $H_R$ describes the kinetic energy for the rotational motion of the linear molecule with the rotational constant $\xi_0$, $\hat {\bf{L}}$ is the angular momentum operator. The rotational eigenstates $|l,m\rangle$ are labeled by the orbital angular quantum number $l$ and its projection $m$ on the laboratory $z$ axis. The second term $H_{ph}$ stands for the kinetic energy of the phonons with the dispersion relation $\omega_{\upsilon}=c k$ (here, the acoustic phonon mode is mainly considered and $c$ is the velocity of phonon mode). ${a_{k \lambda \mu}}$ and $a_{k \lambda \mu}^\dag $ are the annihilation and creation operators for the phonon with the wave vector $k$, which are expressed in the angular momentum representation with $\lambda$ and $\mu$ define, respectively, the phonon angular momentum and its projection on the laboratory-frame $z$ axis; see Refs. 1, 2 and 9 for details.  $H_{R-ph}$ is the interaction term between the rotating molecular cation and the phononic bath with the angular-momentum dependent strength\cite{wjw1,wjw2}
\begin{equation}
{U_{\lambda}(k)}=u_{\lambda}\sqrt{\frac{8\beta k^{3/2}}{2\lambda+1}}\int f_{\lambda}(r)j_{\lambda}(kr)r^2dr,
\end{equation}
where $u_{\lambda}$ and $f_{\lambda}(r)$ define the strength and the interaction potential function in the angular momentum channel $\lambda$, respectively. The parameter $\beta=\sqrt{D^2/(\hbar\rho c^3)}$ describes the acoustic phonon modes stemming from the deformation potential mechanism\cite{wjw29,wjw30,wjw31}, which depends on the deformation potential constant $D$, density $\rho$ and propagating velocity of phonon modes.  $\hat{Y}_{\lambda \mu}(\hat{\theta},\hat{\phi})$ are the spherical harmonics, depending on the angle operators $\hat{\theta}$ and $\hat{\phi}$. $j_{\lambda}(kr)$ is the spherical Bessel function of the first kind and the summation of wave vector satisfies the relation of $\sum_k\equiv\int dk$, the detailed processes have been given in Refs.1 and 2. The octahedral cage of MHP shown in Fig. 1 is approximated by the spherical cage with the effective radius $\Re$, so the interaction potential function $f_{\lambda}(r)$ is assumed satisfying the following form
\begin{equation}
{{ {{{{f_{\lambda} }(r)}}} } } = \left\{\begin{array}{rcl}
&{(\frac{r}{\Re})^{\lambda} },& (r \le \Re)\\
&0,& (r > \Re)\\
\end{array}\right.,
\end{equation}
which can be attributed to the fact that the organic cation rotating freely in the octahedral cage proved by recent experiments\cite{wjw18,wjw19,wjw20}. In principle, this potential distribution function could available to the molecules or impurities rotating in many cage-like structures, such as carbon nanotubes\cite{wjw32} and fullerenes\cite{wjw33}.

In usual, many-body systems describing by Hamiltonian Eq. (1) are very difficult to solve. Analogue to the polaron theory, Schmidt and Lemeshko introduced two canonical transformations to study this problem for the first time\cite{wjw1}. The first canonical transformation allows to transfer the many-body environment degree of freedom into the frame corotating along with the organic cation, which is given by
\begin{equation}
{S_{\rm{1}}} =e^{-i\hat{\phi}\bigotimes\hat{\Lambda}_z}e^{-i\hat{\theta}\bigotimes\hat{\Lambda}_y}e^{-i\hat{\gamma}\bigotimes\hat{\Lambda}_z},
\end{equation}
where $\hat{\phi}$, $\hat{\theta}$, and $\hat{\gamma}$ are the angular operators of the rotational cation in the Hilbert space and $\mathbf{\hat{\Lambda}}=\sum_{k \lambda \mu\nu}a_{k \lambda \mu}^\dag \boldsymbol{\sigma}_{\mu\nu}^{\lambda}{a_{k \lambda \nu}}$ is the collective angular-momentum operator of the many-body phononic bath in Hilbert space with $\boldsymbol{\sigma}^{\lambda}$ stands for the vector of matrices fulfilling the angular-momentum algebra in the angular momentum state $\lambda$. After performing the first transformation, the Hamiltonian of Eq. (1) is converted into\cite{wjw1,wjw2}
\begin{eqnarray}
\hat{H}&=& S_1^{ - 1}{H}{S_1}=\hbar\xi_0 (\hat {\bf{L}}'-\hat{\mathbf{\Lambda}})^{2}+\sum\limits_{k \lambda \mu} {\hbar {\omega_{\upsilon}}a_{k \lambda \mu}^\dag {a_{k \lambda \mu}}}\nonumber\\
&+& \sum\limits_{k\lambda} V_{\lambda}(k) [a_{k\lambda0}^\dag + a_{k\lambda0} ],
\end{eqnarray}
with $V_{\lambda}(k)=U_{\lambda}(k)\sqrt{(2\lambda+1)/(4\pi)}$ and $\hat {\bf{L}}'$ is an abnormal angular momentum operator, because the components of $\hat {\bf{L}}'$ follows abnormal commutation relations in the body fixed frame, differing from the normal operator $\hat {\bf{L}}$ of Eq. (1) in the laboratory frame, the detailed derivations have been given in Ref. 9.

The transformed Hamiltonian is further simplified by introducing the second canonical transformation
\begin{equation}
S_2=\exp{\left[\sum\limits_{k\lambda}\textsf{F}_{k\lambda}^*a_{k\lambda0}^\dag-\textsf{F}_{k\lambda}a_{k\lambda0}\right]},
\end{equation}
where $\textsf{F}_{k\lambda}$ ($\textsf{F}_{k\lambda}^*$) is the variational function. The details of the derivation processes for the second canonical transformation and the variational function are presented in Appendix A. Then, the Hamiltonian is approximated by
\begin{eqnarray}
\widetilde{H}&=&S_2^{-1}\hat{H}S_2\approx \widetilde{H_0}\nonumber\\
 &=&\hbar {\xi _0}{\hat {\bf L}'^2} + \sum\limits_{k\lambda } {} [\hbar {\omega _\upsilon } - \hbar {\xi _0}(\sum\limits_i {{\boldsymbol{\sigma }}_{0i}^\lambda } {\boldsymbol{\sigma }}_{i0}^\lambda )]\nonumber\\
&&\times(a_{k\lambda 0}^\dag  + {\textsf{F}_{k\lambda }})({a_{k\lambda 0}} + \textsf{F}_{k\lambda }^*)-\hbar {\xi _0}( {\hat {\bf{L}}' \cdot \hat {\bf{\Lambda }} + \hat {\bf{\Lambda }} \cdot \hat {\bf{L}}'} )\nonumber\\
 &&+ \sum\limits_{k\lambda } {\left[ {{V_\lambda }(k)( {a_{k\lambda 0}^\dag  + {\textsf{F}_{k\lambda }}} ) + {V_\lambda }(k)( {{a_{k\lambda 0}} + \textsf{F}_{k\lambda }^*} )} \right]}\nonumber\\
  &&+ \hbar {\xi _0}\sum\limits_{k\lambda } {} ({\hat {\bf \Lambda} ^2} + \sum\limits_i {{\boldsymbol{\sigma }}_{0i}^\lambda } {\boldsymbol{\sigma }}_{i0}^\lambda a_{k\lambda 0}^\dag {a_{k\lambda 0}})\nonumber\\
   &&+ \sum\limits_{k\lambda \mu } {} \hbar {\omega _\upsilon }a_{k\lambda \mu }^\dag {a_{k\lambda \mu }} - \sum\limits_{k\lambda } {} \hbar {\omega _\upsilon }a_{k\lambda 0}^\dag {a_{k\lambda 0}},
\end{eqnarray}
with $i=\pm1$. These terms of mainly contribution in $\widetilde{H_0}$ have been reserved. The eigenfunction of angulon can be choose as $\left| \Phi  \right\rangle =|lm\rangle|0_{ph}\rangle$, where $|0_{ph} \rangle$ denotes the vacuum state of phonon, satisfying the relation of $a_{k\lambda0}|0_{ph}\rangle=0$. The expectation values of the eigenenergies of angulon can be obtained via
\begin{eqnarray}
E&=&\hbar {\xi _0}l'(l'+1) + \sum\limits_{k\lambda } {\left[ {\hbar {\omega _\upsilon } + \hbar {\xi _0}\lambda (\lambda  + 1)} \right]{\textsf{F}_{k\lambda }}\textsf{F}_{k\lambda }^*}\nonumber\\
 &&+\sum\limits_{k\lambda } {{V_\lambda }(k)\left( {{\textsf{F}_{k\lambda }} + \textsf{F}_{k\lambda }^*} \right)},
\end{eqnarray}
where $l'$ is the total angular momentum of the system. Minimizing Eq. (8) with respect to $\textsf{F}_{k\lambda}^*$ and $\textsf{F}_{k\lambda}$ (see Appendix A), one can get
\begin{equation}
\textsf{F}_{k\lambda}^*= \textsf{F}_{k\lambda}=-\frac{V_{\lambda}(k)}{\hbar\omega_{\upsilon}+\hbar\xi_0\lambda(\lambda+1)}.
\end{equation}
Substituting Eq. (9) into Eq. (8), the energy of angulon could be expressed as
\begin{equation}
E = \hbar\xi_0l'(l'+1)-\sum\limits_{k\lambda } {\frac{{|{V_\lambda }(k){|^2}}}{{\hbar {\omega _\upsilon } + \hbar {\xi _0}\lambda (\lambda  + 1)}}}.
\end{equation}

\section{Optical absorption of the single angulon}
Based on the Fermi's golden rule, the absorption coefficient $\Gamma(\hbar\Omega)$ for an incident photon with energy $\hbar\Omega$ from the ground state of an angulon is\cite{wjw25,wjw26,wjw27}
\begin{equation}
\Gamma(\hbar\Omega)=\frac{2\pi\hbar\eta}{\varepsilon_\nu B^2}\sum_f|\langle\Phi_0|M|\Phi_f\rangle|^2\delta(E_0+\hbar\Omega-E_f),
\end{equation}
where $\eta$ and $\varepsilon_\nu$ are the permeability and permittivity of the medium, respectively; $\mathbf{B}$ is the magnetic field vector of the incident photon. $M=\gamma_0\textbf{L}\cdot \mathbf{B}$ is the time dependent perturbation term ($\gamma_0=e/2m_0c_0$), describing the orbital magnetic momentum $\gamma_0\textbf{L}$ of the rotational organic cation interacts with the magnetic vector of the incident photon. $\Phi_0$ denotes the ground state of an angulon with the eigenenergy $E_0$. The possible final state is described by $\Phi_f$ with the eigenenergy $E_f$. In fact, the final state includes all possible excited states of the angulon that are so poorly known. Namely, the calculations of absorption coefficient are very difficult in practice.

Several decades ago, Devreese et al.\cite{wjw25} proposed a very simple model without the accuracy to explore the optical absorption of polaron in traditional semiconductors, in which all the excited states of polaron were simplified via two unitary transformations, thus to avoid the accurate information for these final polaron states. Getting an enlightenment from this pioneering work, we expand this model to the optical absorption of angulon. After the strict mathematical deduction presented in Appendix B, the absorption coefficient can be expressed as
\begin{eqnarray}
\Gamma (\hbar \Omega ) &=& \frac{{\pi\hbar \eta \gamma_0^2}}{{{3\varepsilon _\nu}}}\sum\limits_{k\lambda }\hbar {\xi _0}\lambda (\lambda  + 1){\left| {{\mathsf{F}_{k\lambda }}} \right|^2}\nonumber\\
&&\times\delta \left[ {\hbar \Omega  - \hbar {\omega _\upsilon } - \hbar {\xi _0}\lambda (\lambda  + 1)} \right].
\end{eqnarray}
Substituting Eq. (9) into Eq. (12) and converting the summation of the wave vector $k$ into the integral, Eq. (12) becomes
\begin{eqnarray}
\Gamma (\hbar \Omega ) &=& \frac{{4\pi \eta {\gamma_0^2}{c^{3/2}}{\xi _0}\beta }}{{{3\Re^3}{\varepsilon _\nu }}}\nonumber\\
&&\times\frac{{{u_\lambda^2 }\lambda (\lambda  + 1)J_{\lambda  + 3/2}^2\left[ {\Re\left( {\Omega  - {\xi _0}\lambda (\lambda  + 1)} \right)/c} \right]}}{{{{\left( \Omega  \right)}^2}{{\left[ { \Omega  -  {\xi _0}\lambda (\lambda  + 1)} \right]}^{3/2}}}},\nonumber\\
\end{eqnarray}
where $J_{\lambda + 3/2}$ is the Bessel function of the first kind for the quantum state of phonon angular momentum $\lambda$.

In the numerical simulation, the conditions $1.75u_1= 3.06u_2=5.36u_3=u_0$ for the strength parameters for three lowest angular momentum channels are assumed\cite{wjw1,wjw2}. The effective radius $\Re=3.63$ {\AA} and the velocity of acoustic phonon modes $c=2200$ m/s for CH$_3$NH$_3$PbI$_3$ in the cubic phase are selected\cite{wjw16,wjw17}. Taking these parameters into Eq. (13), the optical absorption of an angulon are shown in following figures.
\begin{figure*}
\includegraphics[width=7.0in,keepaspectratio]{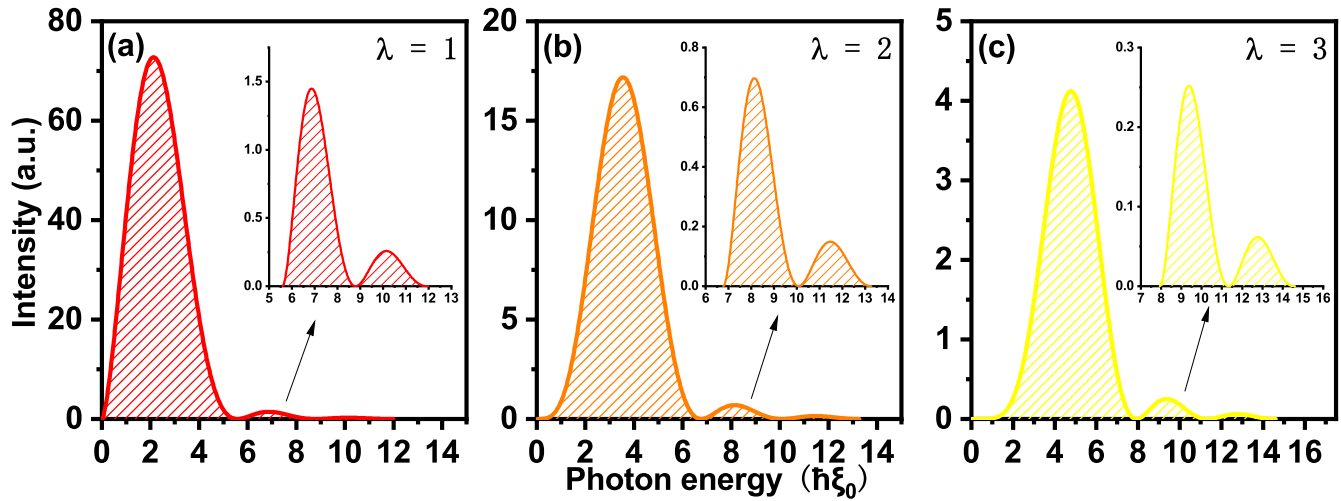}
\caption{\label{compare} The optical absorption of an angulon as a function of the photon energy for different quantum states of phonon angular momentum $\lambda=1$ (a), $\lambda=2$ (b) and $\lambda=3$ (c). The incident photon energy is in units of $\hbar\xi_0$.}
\end{figure*}

Fig. 2 shows the optical absorption of an angulon as a function of the incident photon energy ($\hbar\Omega$) for three different quantum states of phonon angular momentum $\lambda=1, 2, 3$. For $\lambda=1$ in Fig. 2 (a), one can see that the resonance absorption peaks with different densities will be appeared as the energy of incident photon matching the total energies of the phonon and its rotational motion. The similar behaviors are also obtained for $\lambda=2$ and $\lambda=3$ in Figs. (b) and (c), respectively. Obviously, these resonance peaks imply a series of eigenstates of angulon, representing the total angular momentum quantum states of the phonon bath corotating with the organic cation, which suggest that different angulon states could be directly reflected by the absorption spectroscopy. Initially, Schmidt and Lemeshko proposed that the angulon induces a rich rotational fine structure\cite{wjw1,wjw2}, for instance, the rotational Lamb shifts that are very consistent with the previous observations in spectra of the rotating molecules embedded in superfluid helium nanodroplets\cite{wjw34,wjw35,wjw36} and the appearance of two angulon branches due to the isotropic effect of the finite-range potential between rotational particle and phonon bath. Soon afterwards, they proposed that the angulon spectral function for different orbital quantum states shows a discontinuity with the rotational speed of the quantum impurity, implying the behavior of the transfer of one quanta of angular momentum from the phonon bath to the rotating impurity at a special rotational speed\cite{wjw2,wjwz1,wjwj1}. This behavior should be explained by the resonance absorption peak shown in Fig. 2. But only one discontinue point is obtained in their model for the certain orbital angular momentum state. Here, we can predict that there should, in fact, exist other discontinue points denoting the angular momentum transfer with increasing the rotational speed in Ref.2, just like multiple resonance peaks as given in Fig. 2.  Furthermore, they studied the variation of the deformation of phonon density (or the average number of phonon angular momentum) around the instable angulon state and pointed out the variational shape of phonon density could be serve as a fingerprint of the angular-momentum exchange\cite{wjw2,wjwz1,wjwj1}. Yakaboylu et al. found that angulon could induce anomalous screening even in a neutral and weakly polarizable environment and suggested that this angulon can be evaluated just by a single angular-momentum-dependent screening factor related to several observables, such as the effective Rabi frequency, geometric phase and molecular spatial alignment\cite{wjw3}. In addition, Lemeshko also proposed that the angulon effect could be speculated indirectly by renormalization of the effective molecular moments of inertia, which is in good agreement with experimental measurements for a broad range of molecular impurities species\cite{wjw5}. Although so many strategies above mentioned have been proposed for identifying the angulon effect, the optical absorption provides a more direct and effective way to study this problem.

We only discuss one type of rotational motion of the organic cation in the octahedral cage in this paper. Indeed, there are several different types of rotation motions with different rotational inertias in this structure, both of which could be reflected in the resonance absorption with the distinctly rotational constant as shown in Fig. 2.  Hence, this model lays the theoretical foundation to analyze the rotational dynamics of the molecule immersed into the quantum many-body environment by absorption spectroscopy. Usually, MHP materials undergo two structural phase transitions with temperature, such as the crystal phase of CH$_3$NH$_3$PbI$_3$  changes from orthorhombic to tetragonal at temperature $T=150$ K and then to the cubic phase at $T=293$ K\cite{wjw37}. These phase transitions not only have a significant influence on the strength of the electron-phonon interaction and phonon energy, but also modify the distribution of the interaction potential between the rotating cation and the phonon bath. Consequently, different types of rotational motion of cation possessing different the molecular moments of inertia could be reflected directly by the different resonance absorption peaks of angulon. Therefore, the optical absorption of angulon may provide a valuable and direct resolution method to judge the phase transitions in the MHP materials. We hope that these theoretical results can stimulate the progresses of the related experiments.

\section{conclusion}
In summary, the optical absorption of an angulon is studied in MHP based on an improved Devreese-Huybrechts-Lemmens model proposed in traditional semiconductors for polarons. We find that multiple resonance absorption peaks are appeared as the incident photon matches the total energies of acoustic phonon and its rotational motion, where the intensities of optical absorption decrease with quantum states of phonon angular momentum. These theoretical results indicate that (i) optical absorption could be served as an effective and simple method to identify the angulon quasiparticle; (ii) the organic cation rotating in the lattice cage in MHP provides a new platform to study the redistribution of orbital angular momentum in quantum
many-body systems.

\section{Acknowledge}
This work was supported by National Natural Science Foundation of China (Nos. 12174283, 62022081 and 61974099).

\begin{appendix}
\section{The details for the derivation of the variational function}
 Here, we give the details on the derivation of the variational function $\mathsf{F}_{k\lambda}$ ($\mathsf{F}^*_{k\lambda}$).

 In the angular momentum representation, the phonon creation and annihilation operators, $a_{k\lambda \mu }^\dag $ and $a_{k\lambda \mu }$, are defined as irreducible tensors of rank $\lambda$. According to the commutation relations $[{a_{k\lambda \mu }},a_{k\lambda \mu }^\dag ] = 1$ and $[a_{k\lambda \mu }^\dag,a_{k\lambda \mu }^\dag ] =[{a_{k\lambda \mu }},a_{k\lambda \mu } ]=0$, the second canonical transformation ${S_2} = \exp [\sum\nolimits_{k\lambda } {\mathsf{F}_{k\lambda }^*a_{k\lambda 0}^\dag }  - \mathsf{F}_{k\lambda }^{}a_{k\lambda 0}^{}]$ produces a ``shift'' of the phonon operators\cite{wjw2,wjwj1}:
 \begin{eqnarray}
 S_2^{ - 1}{a_{k\lambda 0}}{S_2} = {a_{k\lambda 0}} + \mathsf{F}_{k\lambda }^*,
 \end{eqnarray}
 \begin{eqnarray}
 S_2^{ - 1}a_{k\lambda 0}^\dag {S_2} = a_{k\lambda 0}^\dag  + {\mathsf{F}_{k\lambda }}.
 \end{eqnarray}
 Upon substitution of Eqs. (A1) and (A2), the second term of Eq. (5) transformed by the $S_2$ operator can be of the form
 \begin{eqnarray}
 &&S_2^{ - 1}(\sum\limits_{k\lambda \mu } {\hbar {\omega _\upsilon }} a_{k\lambda \mu }^\dag {a_{k\lambda \mu }}){S_2}\nonumber\\
  &=& S_2^{ - 1}(\sum\limits_{k\lambda \mu  \ne 0} {\hbar {\omega _\upsilon }} a_{k\lambda \mu }^\dag {a_{k\lambda \mu }} + \sum\limits_{k\lambda } {\hbar {\omega _\upsilon }} a_{k\lambda 0}^\dag {a_{k\lambda 0}}){S_2}\nonumber\\
   &=& \sum\limits_{k\lambda \mu  \ne 0} {\hbar {\omega _\upsilon }} a_{k\lambda \mu }^\dag {a_{k\lambda \mu }} + \sum\limits_{k\lambda } {\hbar {\omega _\upsilon }} (a_{k\lambda 0}^\dag  + {\mathsf{F}_{k\lambda }})({a_{k\lambda 0}} + \mathsf{F}_{k\lambda }^*)\nonumber\\
  & = &\sum\limits_{k\lambda } {\hbar {\omega _\upsilon }(} \mathsf{F}_{k\lambda }^*a_{k\lambda 0}^\dag  + {\mathsf{F}_{k\lambda }}{a_{k\lambda 0}} + {\mathsf{F}_{k\lambda }}\mathsf{F}_{k\lambda }^*)\nonumber\\
  &&+\sum\limits_{k\lambda \mu } {\hbar {\omega _\upsilon }} a_{k\lambda \mu }^\dag {a_{k\lambda \mu }}.
 \end{eqnarray}

 Similarly, the third term of Eq. (5) after the second transformation becomes
 \begin{eqnarray}
&&S_2^{ - 1}\left[ {\sum\limits_{k\lambda } {{V_\lambda }(k)( {a_{k\lambda 0}^\dag  + {a_{k\lambda 0}}} )} } \right]{S_2}\nonumber\\
 &=& \sum\limits_{k\lambda } {\left[ {{V_\lambda }(k)\left( {a_{k\lambda 0}^\dag  + {a_{k\lambda 0}} + \mathsf{F}_{k\lambda }^* + {\mathsf{F}_{k\lambda }}} \right)} \right]}.\nonumber\\
 \end{eqnarray}

The transformation of the first term $\hbar\xi_0 (\hat {\bf{L}}'-\hat{\mathbf{\Lambda}})^{2}$ in Eq. (5) turns out to be slightly more cumbersome. The collective angular momentum operator of the phonon bath, $\mathbf{\hat{\Lambda}}=\sum_{k \lambda \mu\nu}a_{k \lambda \mu}^\dag \boldsymbol{\sigma}_{\mu\nu}^{\lambda}{a_{k \lambda \nu}}$, acts in the Hilbert space, where the relations are fulfilled for the phononic states: ${\hat {\mathbf{\Lambda }}^2}\left| {k\lambda \mu } \right\rangle  = \lambda (\lambda  + 1)\left| {k\lambda \mu } \right\rangle$ and ${\hat {{\Lambda }}_z}\left| {k\lambda \mu } \right\rangle  = \mu \left| {k\lambda \mu } \right\rangle$. Consequently, it is transformed by the ${S_2}$ operator in the following way:
  \begin{eqnarray}
&&S_2^{ - 1}\hat {\bf{\Lambda }}{S_2}\nonumber\\
 &=& S_2^{ - 1}( {\sum\limits_{k\lambda \mu \nu } {a_{k\lambda \mu }^\dag {\boldsymbol{\sigma }}_{\mu \nu }^\lambda {a_{k\lambda \nu }}} } ){S_2}\nonumber\\
&=&S_2^{ - 1}(\sum\limits_{k\lambda \mu  \ne 0\nu  \ne 0} {a_{k\lambda \mu }^\dag {\boldsymbol{\sigma }}_{\mu \nu }^\lambda {a_{k\lambda \nu }}}  + \sum\limits_{k\lambda \mu  \ne 0} {a_{k\lambda \mu }^\dag {\boldsymbol{\sigma }}_{\mu 0}^\lambda {a_{k\lambda 0}}}\nonumber\\
&&+ \sum\limits_{k\lambda \nu  \ne 0} {a_{k\lambda 0}^\dag {\boldsymbol{\sigma }}_{0\nu }^\lambda {a_{k\lambda \nu }}}  + \sum\limits_{k\lambda } {a_{k\lambda 0}^\dag {\boldsymbol{\sigma }}_{00}^\lambda {a_{k\lambda 0}}} ){S_2}\nonumber\\
&=& \sum\limits_{k\lambda \mu  \ne 0\nu  \ne 0} {a_{k\lambda \mu }^\dag {\boldsymbol{\sigma }}_{\mu \nu }^\lambda {a_{k\lambda \nu }}}  + \sum\limits_{k\lambda \mu  \ne 0} {a_{k\lambda \mu }^\dag {\boldsymbol{\sigma }}_{\mu 0}^\lambda ({a_{k\lambda 0}}}  + \mathsf{F}_{k\lambda }^*)\nonumber\\
 &&+ \sum\limits_{k\lambda \nu  \ne 0} {(a_{k\lambda 0}^\dag  + {\mathsf{F}_{k\lambda }}){\boldsymbol{\sigma }}_{0\nu }^\lambda {a_{k\lambda \nu }}}\nonumber\\
   &&+ \sum\limits_{k\lambda } {(a_{k\lambda 0}^\dag  + {\mathsf{F}_{k\lambda }}){\boldsymbol{\sigma }}_{00}^\lambda ({a_{k\lambda 0}}}  + \mathsf{F}_{k\lambda }^*)\nonumber\\
   &=& \hat {\bf{\Lambda }} + \sum\limits_{k\lambda \mu } {\mathsf{F}_{k\lambda }^*a_{k\lambda \mu }^\dag {\boldsymbol{\sigma }}_{\mu 0}^\lambda }  + \sum\limits_{k\lambda \nu } {{\mathsf{F}_{k\lambda }}{\boldsymbol{\sigma }}_{0\nu }^\lambda {a_{k\lambda \nu }}}\nonumber\\
     &&+ \sum\limits_{k\lambda } {{\mathsf{F}_{k\lambda }}\mathsf{F}_{k\lambda }^*{\boldsymbol{\sigma }}_{00}^\lambda },
\end{eqnarray}
 where the angular momentum dependent ${{\bf{\sigma }}_{\mu \nu }^\lambda }$ is given by
 \begin{eqnarray}
 {\bf{\sigma }}_{\mu \nu }^\lambda  = \nu{\delta _{\mu,\nu}} \mp \sqrt {\frac{{\lambda (\lambda  + 1) - \nu (\nu  \pm 1)}}{2}} {\delta _{\mu,\nu  \pm 1}}.
 \end{eqnarray}
 Thus, we obtain ${{\bf{\sigma }}_{00}^\lambda }=0$, and rewrite Eq. (A5) as
 \begin{eqnarray}
 S_2^{ - 1}\hat {\bf{\Lambda }}{S_2}=\hat {\bf{\Lambda }} + \sum\limits_{k\lambda \mu } {\mathsf{F}_{k\lambda }^*a_{k\lambda \mu }^\dag {\boldsymbol{\sigma }}_{\mu 0}^\lambda }  + \sum\limits_{k\lambda \nu } {{\mathsf{F}_{k\lambda }}{\boldsymbol{\sigma }}_{0\nu }^\lambda {a_{k\lambda \nu }}}.
 \end{eqnarray}

 In the laboratory frame, the total angular momentum of the phononic bath is defined by its spherical components, $\hat {\bf{\Lambda }} = \{ {\hat {{\Lambda }}_{ - 1}},{\hat {{\Lambda }}_0},{\hat {{\Lambda }}_{+1}}\}$, where
\begin{eqnarray}
{\hat {{\Lambda }}_0} = {\hat {{\Lambda }}_z},
\end{eqnarray}
\begin{eqnarray}
{\hat {{\Lambda }}_{ + 1}} =  - \frac{1}{{\sqrt 2 }}\left( {{{\hat {{\Lambda }}}_x} + i{{\hat {{\Lambda }}}_y}} \right),
\end{eqnarray}
\begin{eqnarray}
{\hat {{\Lambda }}_{ - 1}} = \frac{1}{{\sqrt 2 }}\left( {{{\hat {{\Lambda }}}_x} - i{{\hat {{\Lambda }}}_y}} \right).
\end{eqnarray}
By using the property of Eq. (A7), one can show that the operators Eqs. (A8)-(A10) transform under $S_2$ in the following way:
\begin{eqnarray}
S_2^{ - 1}{\hat {{\Lambda }}_0}{S_2} &=& {\hat {{\Lambda }}_0} + \sum\limits_{k\lambda } {\mathsf{F}_{k\lambda }^*a_{k\lambda 0}^\dag {\bf{\sigma }}_{00}^\lambda }  + \sum\limits_{k\lambda } {{\mathsf{F}_{k\lambda }}{\bf{\sigma }}_{00}^\lambda {a_{k\lambda 0}}}\nonumber\\
  &=& {\hat {{\Lambda }}_0}.
\end{eqnarray}
\begin{eqnarray}
S_2^{ - 1}{\hat {{\Lambda }}_{ + 1}}{S_2} = {\hat {{\Lambda }}_{ + 1}} + \sum\limits_{k\lambda } {\mathsf{F}_{k\lambda }^*a_{k\lambda 1}^\dag {\bf{\sigma }}_{10}^\lambda }  + \sum\limits_{k\lambda } {{\mathsf{F}_{k\lambda }}{\bf{\sigma }}_{0 - 1}^\lambda {a_{k\lambda  - 1}}}.\nonumber\\
\end{eqnarray}
\begin{eqnarray}
S_2^{ - 1}{\hat {{\Lambda }}_{ - 1}}{S_2} = {\hat {{\Lambda }}_{ - 1}} + \sum\limits_{k\lambda } {\mathsf{F}_{k\lambda }^*a_{k\lambda  - 1}^\dag {\bf{\sigma }}_{ - 10}^\lambda }  + \sum\limits_{k\lambda } {{\mathsf{F}_{k\lambda }}{\bf{\sigma }}_{01}^\lambda {a_{k\lambda 1}}}.\nonumber\\
\end{eqnarray}

After angular momentum algebra, we obtain the following expression for the square of the angular momentum in the transformed frame:
\begin{eqnarray}
S_2^{ - 1}{\hat \Lambda ^2}{S_2} = S_2^{ - 1}(\hat {{\Lambda }}_0^2 - {\hat {{\Lambda }}_{ + 1}}{\hat {{\Lambda }}_{ - 1}} - {\hat {{\Lambda }}_{ - 1}}{\hat {{\Lambda }}_{ + 1}}){S_2}.
\end{eqnarray}
For the first term, the expression is written as
\begin{eqnarray}
S_2^{ - 1}\hat {{\Lambda }}_0^2{S_2} = \hat {{\Lambda }}_0^2;
\end{eqnarray}
For the second term, we obtain the expression:
\begin{eqnarray}
&&S_2^{ - 1}{\hat {{\Lambda }}_{ + 1}}{\hat {{\Lambda }}_{ - 1}}{S_2}\nonumber\\
 &=& [{\hat {{\Lambda }}_{ + 1}} + \sum\limits_{k\lambda } {\mathsf{F}_{k\lambda }^*a_{k\lambda 1}^\dag {\bf{\sigma }}_{10}^\lambda }  + \sum\limits_{k\lambda } {{\mathsf{F}_{k\lambda }}{\bf{\sigma }}_{0 - 1}^\lambda {a_{k\lambda  - 1}}} ]\nonumber\\
  &&\times [{\hat {{\Lambda }}_{ - 1}} + \sum\limits_{k'\lambda '} {\mathsf{F}_{k'\lambda '}^*a_{k'\lambda ' - 1}^\dag {\bf{\sigma }}_{ - 10}^{\lambda'} }  + \sum\limits_{k'\lambda '} {{\mathsf{F}_{k'\lambda '}}{\bf{\sigma }}_{01}^{\lambda '}{a_{k'\lambda '1}}} ]\nonumber\\
  &=& {\hat {{\Lambda }}_{ + 1}}{\hat {{\Lambda }}_{ - 1}} + \sum\limits_{k'\lambda '} {\mathsf{F}_{k'\lambda '}^*{{\hat {{\Lambda }}}_{ + 1}}a_{k'\lambda ' - 1}^\dag {\bf{\sigma }}_{ - 10}^{\lambda'} }\nonumber\\
    &&+ \sum\limits_{k'\lambda '} {{\mathsf{F}_{k'\lambda '}}{{\hat {{\Lambda }}}_{ + 1}}{\bf{\sigma }}_{01}^{\lambda '}{a_{k'\lambda '1}}}  + \sum\limits_{k\lambda } {\mathsf{F}_{k\lambda }^*a_{k\lambda 1}^\dag {\bf{\sigma }}_{10}^\lambda }{{\hat {{\Lambda }}}_{ - 1}}\nonumber\\
    &&+ \sum\limits_{k\lambda } {{\mathsf{F}_{k\lambda }}{\bf{\sigma }}_{0 - 1}^\lambda {a_{k\lambda  - 1}}}{{\hat {{\Lambda }}}_{ - 1}}\nonumber\\
      &&+ \sum\limits_{kk'\lambda \lambda '} {\mathsf{F}_{k\lambda }^*\mathsf{F}_{k'\lambda '}^*a_{k\lambda 1}^\dag {\bf{\sigma }}_{10}^\lambda } a_{k'\lambda ' - 1}^\dag {\bf{\sigma }}_{ - 10}^{\lambda'}\nonumber\\
        &&+ \sum\limits_{kk'\lambda \lambda '} {{\mathsf{F}_{k'\lambda '}}\mathsf{F}_{k\lambda }^*a_{k\lambda 1}^\dag {\bf{\sigma }}_{10}^\lambda } {\bf{\sigma }}_{01}^{\lambda '}{a_{k'\lambda '1}}\nonumber\\
           &&+ \sum\limits_{kk'\lambda \lambda '} {{\mathsf{F}_{k\lambda }}\mathsf{F}_{k'\lambda '}^*{\bf{\sigma }}_{0 - 1}^\lambda {a_{k\lambda  - 1}}a_{k'\lambda ' - 1}^\dag {\bf{\sigma }}_{ - 10}^{\lambda'} }\nonumber\\
             &&+ \sum\limits_{kk'\lambda \lambda '} {{\mathsf{F}_{k\lambda }}{\mathsf{F}_{k'\lambda '}}{\bf{\sigma }}_{0 - 1}^\lambda {a_{k\lambda  - 1}}} {\bf{\sigma }}_{01}^{\lambda '}{a_{k'\lambda '1}}.
\end{eqnarray}
It is remarkable that (i) merely extracting the term ( $k=k'$ and $\lambda=\lambda'$) from the expression eliminates the divergency from the problem; (ii) the terms with zero value and with double phonons should be removed. Therefore, the above expression is simplified as:
\begin{eqnarray}
S_2^{ - 1}{\hat {{\Lambda }}_{ + 1}}{\hat {{\Lambda }}_{ - 1}}{S_2}
&=& {\hat {{\Lambda }}_{ + 1}}{\hat {{\Lambda }}_{ - 1}}+ \sum\limits_{k\lambda } {} (\mathsf{F}_{k\lambda }^*{\hat {{\Lambda }}_{ + 1}}a_{k\lambda  - 1}^\dag {\bf{\sigma }}_{ - 10}^\lambda\nonumber\\
  &&+ {\mathsf{F}_{k\lambda }}{\bf{\sigma }}_{0 - 1}^\lambda {a_{k\lambda  - 1}}{\hat {{\Lambda }}_{ - 1}}\nonumber\\
   &&+ {\mathsf{F}_{k\lambda }}\mathsf{F}_{k\lambda }^*{\bf{\sigma }}_{0 - 1}^\lambda {a_{k\lambda  - 1}}a_{k\lambda  - 1}^\dag {\bf{\sigma }}_{ - 10}^\lambda )\nonumber\\
   &=& {\hat {{\Lambda }}_{ + 1}}{\hat {{\Lambda }}_{ - 1}}\nonumber\\
    &&+ \sum\limits_{k\lambda } {} (\mathsf{F}_{k\lambda }^*a_{k\lambda 0}^\dag {\bf{\sigma }}_{0 - 1}^\lambda {a_{k\lambda  - 1}}a_{k\lambda  - 1}^\dag {\bf{\sigma }}_{ - 10}^\lambda\nonumber\\
      &&+ {\mathsf{F}_{k\lambda }}{\bf{\sigma }}_{0 - 1}^\lambda {a_{k\lambda  - 1}}a_{k\lambda  - 1}^\dag {\bf{\sigma }}_{ - 10}^\lambda {a_{k\lambda 0}}\nonumber\\
       &&+ {\mathsf{F}_{k\lambda }}\mathsf{F}_{k\lambda }^*{\bf{\sigma }}_{0 - 1}^\lambda {a_{k\lambda  - 1}}a_{k\lambda  - 1}^\dag {\bf{\sigma }}_{ - 10}^\lambda )\nonumber\\
  &=& {\hat {{\Lambda }}_{ + 1}}{\hat {{\Lambda }}_{ - 1}} + \sum\limits_{k\lambda } {} (\mathsf{F}_{k\lambda }^*a_{k\lambda 0}^\dag {\bf{\sigma }}_{0 - 1}^\lambda {\bf{\sigma }}_{ - 10}^\lambda\nonumber\\
  &&  + {\mathsf{F}_{k\lambda }}{\bf{\sigma }}_{0 - 1}^\lambda {\bf{\sigma }}_{ - 10}^\lambda {a_{k\lambda 0}} + {\mathsf{F}_{k\lambda }}\mathsf{F}_{k\lambda }^*{\bf{\sigma }}_{0 - 1}^\lambda {\bf{\sigma }}_{ - 10}^\lambda ).\nonumber\\
\end{eqnarray}
For the third term, similarly, we obtain the expression:
\begin{eqnarray}
S_2^{ - 1}{\hat {{\Lambda }}_{ - 1}}{\hat {{\Lambda }}_{ + 1}}{S_2} &=& {\hat {{\Lambda }}_{ - 1}}{\hat {{\Lambda }}_{ + 1}} + \sum\limits_{k\lambda } {} ({\mathsf{F}_{k\lambda }}{\bf{\sigma }}_{01}^\lambda {\bf{\sigma }}_{10}^\lambda {a_{k\lambda 0}}\nonumber\\
 &&+ \mathsf{F}_{k\lambda }^*a_{k\lambda 0}^\dag {\bf{\sigma }}_{01}^\lambda {\bf{\sigma }}_{10}^\lambda  + {\mathsf{F}_{k\lambda }}\mathsf{F}_{k\lambda }^*{\bf{\sigma }}_{01}^\lambda {\bf{\sigma }}_{10}^\lambda ).\nonumber\\
\end{eqnarray}

After the second canonical transformation $S_2$, the total Hamiltonian converts into
\begin{eqnarray}
\widetilde{H} = S_2^{ - 1}\hat H{S_2} &=&S_2^{ - 1}[\hbar\xi_0 (\hat {\bf{L}}'-\hat{\mathbf{\Lambda}})^{2} + \sum\limits_{k\lambda \mu } {\hbar {\omega _\upsilon }} a_{k\lambda \mu }^\dag {a_{k\lambda \mu }}\nonumber\\
 &&+ \sum\limits_{k\lambda } {{V_\lambda }(k)( {a_{k\lambda 0}^\dag  + {a_{k\lambda 0}}} )}]{S_2}\nonumber\\
  &\approx& {\widetilde {H_0}},
\end{eqnarray}
\begin{eqnarray}
{\widetilde {H_0}}{\rm{ }} &=& \hbar {\xi _0}{\hat {\bf L}'^2} + \sum\limits_{k\lambda } {} [\hbar {\omega _\upsilon } - \hbar {\xi _0}(\sum\limits_i {{\boldsymbol{\sigma }}_{0i}^\lambda } {\boldsymbol{\sigma }}_{i0}^\lambda )]\nonumber\\
&&\times(a_{k\lambda 0}^\dag  + {\mathsf{F}_{k\lambda }})({a_{k\lambda 0}} + \mathsf{F}_{k\lambda }^*)-\hbar {\xi _0}( {\hat {\bf{L}}' \cdot \hat {\bf{\Lambda }} + \hat {\bf{\Lambda }} \cdot \hat {\bf{L}}'} )\nonumber\\
 &&+ \sum\limits_{k\lambda } {\left[ {{V_\lambda }(k)( {a_{k\lambda 0}^\dag  + {\mathsf{F}_{k\lambda }}} ) + {V_\lambda }(k)( {{a_{k\lambda 0}} + \mathsf{F}_{k\lambda }^*} )} \right]}\nonumber\\
  &&+ \hbar {\xi _0}\sum\limits_{k\lambda } {} ({\hat {\bf \Lambda} ^2} + \sum\limits_i {{\boldsymbol{\sigma }}_{0i}^\lambda } {\boldsymbol{\sigma }}_{i0}^\lambda a_{k\lambda 0}^\dag {a_{k\lambda 0}})\nonumber\\
   &&+ \sum\limits_{k\lambda \mu } {} \hbar {\omega _\upsilon }a_{k\lambda \mu }^\dag {a_{k\lambda \mu }} - \sum\limits_{k\lambda } {} \hbar {\omega _\upsilon }a_{k\lambda 0}^\dag {a_{k\lambda 0}},
\end{eqnarray}
where $i=\pm1$. The eigenfunction of angulon can be chosen as $\left| \Phi  \right\rangle  = \left| {lm}  \right\rangle \left| {{0_{ph}}} \right\rangle$, where the vacuum state of phonons satisfies the relations of ${a_{k\lambda \mu }}\left| {{0_{ph}}} \right\rangle  = 0$ and $a_{k\lambda \mu }^\dag \left| {{0_{ph}}} \right\rangle  = \left| {k\lambda \mu } \right\rangle$. The expectation value of the energy of the angulon can be obtained via
\begin{eqnarray}
E &=& \left\langle \Phi  \right|\widetilde H\left| \Phi  \right\rangle \nonumber\\
  &=& \hbar {\xi _0}l'(l'+1) + \sum\limits_{k\lambda } {\hbar {\omega _\upsilon }{\mathsf{F}_{k\lambda }}\mathsf{F}_{k\lambda }^*}\nonumber\\
    &&+ \sum\limits_{k\lambda } {\hbar {\xi _0}( - {\mathsf{F}_{k\lambda }}\mathsf{F}_{k\lambda }^*{\bf{\sigma }}_{01}^\lambda {\bf{\sigma }}_{10}^\lambda  - {\mathsf{F}_{k\lambda }}\mathsf{F}_{k\lambda }^*{\bf{\sigma }}_{0 - 1}^\lambda {\bf{\sigma }}_{ - 10}^\lambda )}\nonumber\\
      &&+ \sum\limits_{k\lambda } {\left[ {{V_\lambda }(k)\left( {{\mathsf{F}_{k\lambda }} + \mathsf{F}_{k\lambda }^*} \right)} \right]},
\end{eqnarray}
where $l'$ is the total angular momentum of the system, ${\bf{\sigma }}_{01}^\lambda  = {\bf{\sigma }}_{ - 10}^\lambda  = {[\lambda (\lambda  + 1)/2]^{1/2}}$ and ${\bf{\sigma }}_{10}^\lambda  = {\bf{\sigma }}_{0 - 1}^\lambda  =  - {[\lambda (\lambda  + 1)/2]^{1/2}}$. Minimizing Eq. (A21) with respect to $\mathsf{F}_{k\lambda }$ and $\mathsf{F}_{k\lambda }^*$, one can get
\begin{eqnarray}
{\mathsf{F}_{k\lambda }} =  - \frac{{{V_\lambda }(k)}}{{\hbar {\omega _\upsilon } + \hbar {\xi _0}\lambda (\lambda  + 1)}},
\end{eqnarray}
\begin{eqnarray}
\mathsf{F}_{k\lambda }^* =  - \frac{{{V_\lambda }(k)}}{{\hbar {\omega _\upsilon } + \hbar {\xi _0}\lambda (\lambda  + 1)}}.
\end{eqnarray}
\section{Derivation for the optical absorption of an angulon}
The representation of the $\delta$-function can be written as
\begin{eqnarray}
\delta (x) = \frac{1}{\pi }{\mathop{\rm Re}\nolimits} \int_{ - \infty }^0 {dt\exp [ - i(x + i\varepsilon )t]}.
\end{eqnarray}
We converted the absorption coefficient $\Gamma(\hbar\Omega)$ into\cite{wjw25,wjw26,wjw27}
\begin{eqnarray}
\Gamma \left( {\hbar \Omega } \right)&=& \frac{{2 \hbar \eta }}{{{\varepsilon _\nu }{B^2}}}{\mathop{\rm Re}\nolimits}\sum\limits_f {\int_{ - \infty }^0 {dt\left\langle {{\Phi _0}\left| {{ M}\left| {{\Phi _f}} \right\rangle } \right.} \right.} } \left\langle {{\Phi _f}\left| {{ M}\left| {{\Phi _0}} \right\rangle } \right.} \right.\nonumber\\
&&\times{e^{ - i({E_0} + \hbar \Omega  - {E_f} + i\varepsilon )t}}\nonumber\\
&=& \frac{{2\hbar \eta }}{{{\varepsilon _\nu }{B^2}}}{\mathop{\rm Re}\nolimits} \sum\limits_f {\int_{ - \infty }^0 {dt\exp [ - i(\hbar \Omega  + i\varepsilon )t]\left\langle {{\Phi _0}\left| {{ M}\left| {{\Phi _f}} \right\rangle } \right.} \right.} }\nonumber\\
 &&\times\left\langle {{\Phi _f}\left| {{e^{iHt}}{ M}{e^{ - iHt}}\left| {{\Phi _0}} \right\rangle } \right.} \right.\nonumber\\
&=&\frac{{2\hbar \eta }}{{{\varepsilon _\nu }{B^2}}}{\mathop{\rm Re}\nolimits} \int_{ - \infty }^0 {dt\exp [ - i(\hbar \Omega  + i\varepsilon )t]}\nonumber\\
 &&\times\left\langle {{\Phi _0}\left| {{ M}(0){ M}(t)\left| {{\Phi _0}} \right\rangle } \right.} \right.\nonumber\\
&= &\frac{{2\hbar \eta \gamma_0^2}}{{{\varepsilon _\nu }{B^2}}}{\mathop{\rm Re}\nolimits} \int_{ - \infty }^0 {dt\exp [ - i(\hbar \Omega  + i\varepsilon )t]}\nonumber\\
 &&\times\left\langle {{\Phi _0}\left| {({\bf{L}} \cdot {\bf{B}})({\bf{L}}(t) \cdot {\bf{B}})\left| {{\Phi _0}} \right\rangle } \right.} \right.\nonumber\\
\end{eqnarray}

After performing the first and second canonical transformation, respectively, the matrix in Eq. (B2) becomes
\begin{eqnarray}
&&\left\langle {{\Phi _0}\left| {({\bf{L}} \cdot {\bf{B}})({\bf{L}}(t) \cdot {\bf{B}})\left| {{\Phi _0}} \right\rangle } \right.} \right.\nonumber\\
&&= \left\langle {0\left| {S_2^{ - 1}S_1^{ - 1}{\bf{L}} \cdot {\bf{B}}{S_1}S_1^{ - 1}{\bf{L}}(t) \cdot {\bf{B}}{S_1}{S_2}\left| 0 \right\rangle } \right.} \right..
\end{eqnarray}
When we set the abnormal angular operator $\hat L'=0$,  $\hat L_i$ is set to 0 for the ground state of the rotational cation. The first canonical transformations of the dot products ${\bf{{\rm B}}} \cdot {\bf{L}}(t)$ and ${\bf{{\rm B}}} \cdot {\bf{L}}(0)$ are listed below:
\begin{eqnarray}
S_1^{ - 1}{\bf{{\bf B}}} \cdot {\bf{L}}{S_1} &=&  - {\bf{{\bf B}}} \cdot \sum\limits_{k\lambda \mu \nu iq} {D_{iq}^{1*}{\bf \Lambda _q}}\nonumber\\  &=&  - {\bf{{\bf B}}} \cdot \sum\limits_{k\lambda \mu \nu iq} {D_{iq}^{1*}\boldsymbol \sigma _{\mu \nu }^\lambda a_{k\lambda \mu }^\dag } {a_{k\lambda \nu }},
\end{eqnarray}

\begin{eqnarray}
S_1^{ - 1}{\bf{B}} \cdot {\bf{L}}(t){S_1}{\rm{ }} &=& {\bf{B}} \cdot S_1^{ - 1}{e^{iHt}}{\bf{L}}{e^{ - iHt}}{S_1}\nonumber\\
 &=& {\bf{B}} \cdot S_1^{ - 1}{e^{iHt}}{S_1}S_1^{ - 1}{\bf{L}}{S_1}S_1^{ - 1}{e^{ - iHt}}{S_1}\nonumber\\
  &=& {\bf{B}} \cdot {e^{i\hat Ht}}S_1^{ - 1}{\bf{L}}{S_1}{e^{ - i\hat Ht}}\nonumber\\
   &= & - {\bf{B}} \cdot \sum\limits_{k\lambda \mu \nu iq} {D_{iq}^{1*}\boldsymbol\sigma _{\mu \nu }^\lambda a_{k\lambda \mu }^\dag (t)} {a_{k\lambda \nu }}(t).\nonumber\\
\end{eqnarray}
$D_{iq}^1$ are the Wigner rotation matrices. With the help of transformations Eq. (B4) and Eq. (B5), Eq. (B3) becomes
\begin{widetext}
\begin{eqnarray}
&&\left\langle {{\Phi _0}\left| {({\bf{L}} \cdot {\bf{B}})({\bf{L}}(t) \cdot {\bf{B}})\left| {{\Phi _0}} \right\rangle } \right.} \right.\nonumber\\
&=& \left\langle {0\left| {S_2^{ - 1}\left( {{\bf{B}} \cdot \sum\limits_{k\lambda \mu \nu iq} {D_{iq}^{1*}\boldsymbol\sigma _{\mu \nu }^\lambda a_{k\lambda \mu }^\dag } {a_{k\lambda \nu }}} \right)\left( {{\bf{B}} \cdot \sum\limits_{k\lambda \mu \nu iq} {D_{iq}^{1*}\boldsymbol\sigma _{\mu \nu }^\lambda a_{k\lambda \mu }^\dag \left( t \right)} {a_{k\lambda \nu }}\left( t \right)} \right){S_2}} \right|0} \right\rangle\nonumber\\
&=&\left\langle {0\left| {S_2^{ - 1}\left( {{\bf{B}} \cdot \sum\limits_{k\lambda \mu \nu iq} {D_{iq}^{1*}\boldsymbol\sigma _{\mu \nu }^\lambda a_{k\lambda \mu }^\dag } {a_{k\lambda \nu }}} \right){S_2}S_2^{ - 1}{e^{i\widehat Ht}}{S_2}S_2^{ - 1}\left( {{\bf{B}} \cdot \sum\limits_{k\lambda \mu \nu iq} {D_{iq}^{1*}\boldsymbol\sigma _{\mu \nu }^\lambda a_{k\lambda \mu }^\dag } {a_{k\lambda \nu }}} \right){S_2}S_2^{ - 1}{e^{ - i\widehat Ht}}{S_2}} \right|0} \right\rangle\nonumber\\
&=&\left\langle {0\left| {{B^2}\left( {\sum\limits_{k\lambda iq\nu } {D_{iq}^{1*}\sigma _{0\nu }^\lambda {a_{k\lambda \nu }}} {\mathsf{F}_{k\lambda }}} \right){e^{i{{\widetilde H}_0}t}}\left( {\sum\limits_{k\lambda iq\mu } {D_{iq}^{1*}\sigma _{\mu 0}^\lambda a_{k\lambda \mu }^\dag } \mathsf{F}_{k\lambda }^*} \right){e^{ - i{{\widetilde H}_0}t}}} \right|0} \right\rangle,
\end{eqnarray}
\end{widetext}
where the average spatial summations are performed for the polarization orientations of incident light respecting to the orbital magnetic momentum $\bf{B}\cdot \boldsymbol{\sigma}_{\mu\nu}^{\lambda}$ and used in the following derivations. $\mathsf{F}_{k\lambda }^*$ and $\mathsf{F}_{k\lambda }$ are the variational functions obtained in Eqs. (A22) and (A23). From the equation of motion for ${a_{k\lambda \mu }^\dag }$:
\begin{eqnarray}
\frac{{da_{k\lambda \mu }^\dag (t)}}{{dt}} = i\left[ {{{\widetilde H}_0},a_{k\lambda \mu }^\dag } \right] = i\left( {\hbar {\omega _\upsilon } + {\xi _0}\hbar \lambda (\lambda  + 1)} \right)a_{k\lambda \mu }^\dag,\nonumber\\
\end{eqnarray}
we get
\begin{eqnarray}
{e^{i{{\widetilde H}_0}t}}a_{k\lambda \mu }^\dag {e^{ - i{{\widetilde H}_0}t}} = a_{k\lambda \mu }^\dag \exp [i\left( {\hbar {\omega _\upsilon } + \hbar {\xi _0}\lambda (\lambda  + 1)} \right)t].\nonumber\\
\end{eqnarray}
Thus, the matrix element converts into
\begin{eqnarray}
&&\left\langle {{\Phi _0}\left| {({\bf{L}} \cdot {\bf{B}})({\bf{L}}(t) \cdot {\bf{B}})\left| {{\Phi _0}} \right\rangle } \right.} \right.\nonumber\\
 &=& \left\langle {0\left| {{B^2}\sum\limits_{k\lambda \mu } {D_{0 - \mu }^{1*}\sigma _{0\mu }^\lambda } D_{0\mu }^{1*}\sigma _{\mu 0}^\lambda {\mathsf{F}_{k\lambda }}\mathsf{F}_{k\lambda }^*} \right|0} \right\rangle\nonumber\\
  &&\times\exp [i\left( {\hbar {\omega _\upsilon } + \hbar {\xi _0}\lambda (\lambda  + 1)} \right)t].
\end{eqnarray}
The absorption coefficient is easily calculated from the Eq. (B9)
\begin{eqnarray}
\Gamma (\hbar \Omega ) &=& \frac{{\pi\hbar \eta\gamma_0^2 }}{{{3\varepsilon _\nu }{B^2}}}\sum\limits_{k\lambda } {{B^2}}\hbar {\xi _0}\lambda (\lambda  + 1){\left| {{\mathsf{F}_{k\lambda }}} \right|^2}\nonumber\\
&&\times\delta \left[ {\hbar \Omega  - \hbar {\omega _\upsilon } - \hbar {\xi _0}\lambda (\lambda  + 1)} \right].
\end{eqnarray}

In the present paper, the dispersion relation ${\omega _\upsilon } = ck$ for the acoustic phonon modes is mainly considered. Then, Eq.(2) becomes
\begin{eqnarray}
{V_\lambda }\left( k \right) &=& \sqrt {\frac{{2\lambda  + 1}}{{4\pi }}} {U_\lambda }\left( k \right)\nonumber\\
 &=& {u_\lambda }\sqrt {\frac{{2\beta {k^{3/2}}}}{\pi }} \int_0^{\Re} {{f_\lambda }\left( r \right){j_\lambda }\left( {kr} \right){r^2}dr}\nonumber\\
&=& {u_\lambda }\sqrt {\beta {k^{3/2}}}  \frac{{{J_{\lambda  + 3/2}}\left( {k{\Re}} \right)}}{{{{\left( {k{\Re}} \right)}^{3/2}}}}.
\end{eqnarray}

Finally, substituting Eq. (B11) into Eq. (B10), the absorption coefficient can be expressed as
\begin{eqnarray}
\Gamma (\hbar \Omega ) &=& \frac{{4\pi \eta {\gamma_0^2}{c^{3/2}}{\xi _0}\beta }}{{{3\Re^3}{\varepsilon _\nu }}}\nonumber\\
&&\times\frac{{{u_\lambda^2 }\lambda (\lambda  + 1)J_{\lambda  + 3/2}^2\left[ {\Re\left( {\Omega  - {\xi _0}\lambda (\lambda  + 1)} \right)/c} \right]}}{{{{\left( \Omega  \right)}^2}{{\left[ { \Omega  -  {\xi _0}\lambda (\lambda  + 1)} \right]}^{3/2}}}}.\nonumber\\
\end{eqnarray}
\end{appendix}

\end{document}